\documentclass[conference]{IEEEtran}
\IEEEoverridecommandlockouts
\usepackage{cite}
\usepackage{amsmath,amssymb,amsfonts}
\usepackage{algorithmic}
\usepackage{graphicx}
\usepackage{textcomp}
\usepackage{xcolor}
\usepackage[acronym]{glossaries}
\usepackage{hyperref}
\usepackage{cleveref}
\usepackage{multirow}
\usepackage[table,xcdraw]{xcolor}
\usepackage{soul}
\def\BibTeX{{\rm B\kern-.05em{\sc i\kern-.025em b}\kern-.08em
    T\kern-.1667em\lower.7ex\hbox{E}\kern-.125emX}}

\begin{document}
\newacronym{gpu}{GPU}{\,Graphics\,Processing\,Unit}
\newacronym{cnn}{CNN}{\,Convolutional\,Neural\,Network}
\newacronym{sde}{SDE}{\,Silent\,Data\,Error}
\newacronym{aii}{AII}{\,Active\,Issued\,Instructions}
\newacronym{aei}{AEI}{\,Active\,Executed\,Instructions}
\newacronym{nbti}{NBTI}{\,Negative\,Bias\,Temperature\,Instability}
\newacronym{sm}{SM}{\,Streaming\,Multiprocessor}
\newacronym{hpc}{HPC}{\,High-Performance\,Computing}
\newacronym{cf}{CF}{\,SM\,Clock\,Frequency}
\newacronym{temp}{Temp}{\,Temperature}
\newacronym{pow}{Pow}{\,Power\,consumption}
\newacronym{energy}{E}{\,Energy\,Consumption}
\newacronym{cupti}{CUPTI}{\,CUDA\,Profiling\,Tools\,Interface}
\newacronym{nvml}{NVML}{\,NVIDIA\,Management\,Library}
\newacronym{ncu}{NCU}{\,NVIDIA\,Compute\,Utility}
\newacronym{fit}{FIT}{\,Failure\,In\,Time}
\newacronym{dvfs}{DVFS}{\,Dynamic\,Voltage\,and\,Frequency\,Scaling}

\title{
% Estimating the GPU Stress of Real Applications Through Telemetry and Performance Counters\\
GPU Under Pressure: Estimating Application's Stress via Telemetry and Performance Counters\\
% {\color{blue} From Data to Stress: GPU Load Estimation Using Performance Telemetry}\\
% {\color{green}Performance Counters in Action: Measuring Real-Application GPU Stress}
\vspace{-3mm}}

\author{%

  \IEEEauthorblockN{%
    Giuseppe Esposito\IEEEauthorrefmark{1}, 
    Juan-David Guerrero-Balaguera\IEEEauthorrefmark{1},
    Josie E. Rodriguez Condia\IEEEauthorrefmark{1}, 
    Matteo Sonza Reorda\IEEEauthorrefmark{1},\\
    Marco Barbiero\IEEEauthorrefmark{2}, 
    Rossella Fortuna\IEEEauthorrefmark{2}
  }
  \IEEEauthorblockA{\IEEEauthorrefmark{1}%
    Politecnico di Torino, Dept. of Control and Computer Engineering (DAUIN), Turin, Italy\\
    \{giuseppe.esposito, juan.guerrero, josie.rodriguez, matteo.sonzareorda\}@polito.it
  }
  \IEEEauthorblockA{\IEEEauthorrefmark{2}%
    AI Delivery Factory, Intesa Sanpaolo S.p.A., Turin, Italy\    
    \{marco.barbiero, rossella.fortuna\}@intesasanpaolo.com
  }
%  \vspace{-10mm} % such spaces are not required. It means that you need to compact the text
}
% \author{Anonymous authors}
\newcommand\todo[1]{\textcolor{red}{#1}}

\maketitle

\begin{abstract} 
    Graphics Processing Units (GPUs) are specialized accelerators in data centers and high-performance computing (HPC) systems, enabling the fast execution of compute-intensive applications, such as Convolutional Neural Networks (CNNs). However, sustained workloads can impose significant stress on GPU components, raising reliability concerns due to potential faults that corrupt the intermediate application computations, leading to incorrect results. Estimating the stress induced by an application is thus crucial to predict reliability (with\,special\,emphasis\,on\,aging\,effects). In this work, we combine online telemetry parameters and hardware performance counters to assess GPU stress induced by different applications. 
    The experimental results indicate the stress induced by a parallel workload can be estimated by combining telemetry data and Performance Counters that reveal the efficiency in the resource usage of the target workload. For this purpose the selected performance counters focus on measuring the \textit{i)} throughput, \textit{ii)} amount of issued instructions and \textit{iii)} stall events.

\end{abstract}

\begin{IEEEkeywords}
Graphics Processing Units (GPUs), Application Stress, Performance Counters, Artificial Intelligence (AI), Reliability
\vspace{-3mm}
\end{IEEEkeywords}

\section{Introduction}
\label{sec:intro}

In recent years, \,Graphics\,Processing\,Units\,(GPUs) have become a cornerstone of \,High-Performance\,Computing\,(HPC) systems and modern data centers, powering a wide array of applications ranging from medical imaging to scientific simulations and the training of large-scale Neural Networks \cite{assiroj2018high}. The architectural design of GPUs—featuring hundreds to thousands of cores organized in clusters—enables the parallel execution of thousands of software threads. % , according to the Single Instruction Multiple Data which allows to execute multiple threads at the same time. % \todo{Is it important to mention details about the GPU internal architecture and operation?}

While this operational model provides significant performance improvements, it also raises concerns about the long-term reliability of the hardware. During periods of sustained high device utilization, the high transistor density causes the silicon device to heat up internally, potentially leading to premature aging and physical degradation.
Sustained high utilization and harsh operating conditions can accelerate physical degradation mechanisms. % \todo{It is true that temperature is the main factor that induces aeging, however in this paragraph it would be good to indicate the reasons that can make such temperature increase, e.g., self heating of silicon devices due to high transistor densities is one cause, exacerbated by the prolonged execution of parallel workloads}. 
Among the most critical are \,Negative\,Bias\,Temperature\,Instability (NBTI) and electromigration. 
% NBTI increases the threshold voltage in PMOS transistors under negative gate bias, which reduces operational frequency and overall performance. Electromigration involves high current densities, potentially causing open or short circuits. 
Over time, these phenomena can produce \,Silent\,Data\,Errors (SDEs) that propagate through application execution, inducing system failures, thereby jeopardizing the system reliability~\cite{bao2021critical, lee2024damage}. Moreover, intensive workloads executed on GPUs can expose them to significant physical stress, such as self-heating, which increases the likelihood of hardware faults when used for extended periods~\cite{duan2024efficient}. For instance, premature aging of the GPU in the Titan supercomputing system was found to occur, on average, every 2.8 years, necessitating the replacement of 9,500 GPUs \cite{ostrouchov2020gpu}.

Detecting faults in GPUs that cause SDEs is challenging because they usually occur during in-field operations under variable environmental and workload conditions, like high temperatures and heavy computational demands. To address this, in-field functional test strategies have been implemented using highly intensive workloads. These tests simulate real operational conditions, pushing the devices to their limits to activate and detect potential faults. Some examples of functional in-field testing through induced stress are OpenDCDiag and GPU-burn. The former aims to detect SDE on Intel CPUs during self-tests under varying workloads \cite{macieira2024silent}, whereas GPU-burn applies intense stress, aiming to trigger and detect hardware faults~\cite{defour2013gpuburn}.

Although the use of stress-induced testing solutions has demonstrated effectiveness in enhancing fault detection capabilities, they must be applied periodically to detect possible faults way before they produce catastrophic results \cite{tuncer2017diagnosing}. 
Testing GPUs regularly with GPU-burn can be problematic. Intense testing can significantly shorten their lifespan due to accelerated degradation. Conversely, infrequent testing, limited to maintenance schedules, risks overlooking faults that could lead to downtime, ultimately affecting costs, technical performance, and resource management.
% In the case of GPUs, the adoption of periodic testing using GPU-burn is challenging due to several factors. Firstly,  highly intensive testing procedures applied frequently may accelerate the GPUs' degradation heavily, reducing their lifespan. On the other hand, the application of sporadic testing or during the scheduled maintenance periods only might lead to any fault appearing during the operational phase of the GPU and consequently cause downtime events, meaning economical, technical, and resource impacts. Therefore, determining an appropriate test frequency is crucial to balancing diagnostic effectiveness against system availability and long-term durability. 
Effective stress level estimation for devices, e.g., GPUs, is crucial for maintaining this balance. Devices that frequently handle heavy workloads need regular checks, while those under lighter conditions can have longer intervals between tests without losing reliability \cite{prakash2021transistor}.
%\todo{[unclear if this statement comes from the author or from reference studies, a strong reference can contribute]}. 
Consequently, it is essential to characterize the stress induced by the workloads imposed on GPU to determine the frequency at which in-field functional testing procedures should be applied.

\begin{figure}[tb]
    \centering
    \includegraphics[width=0.9\linewidth]{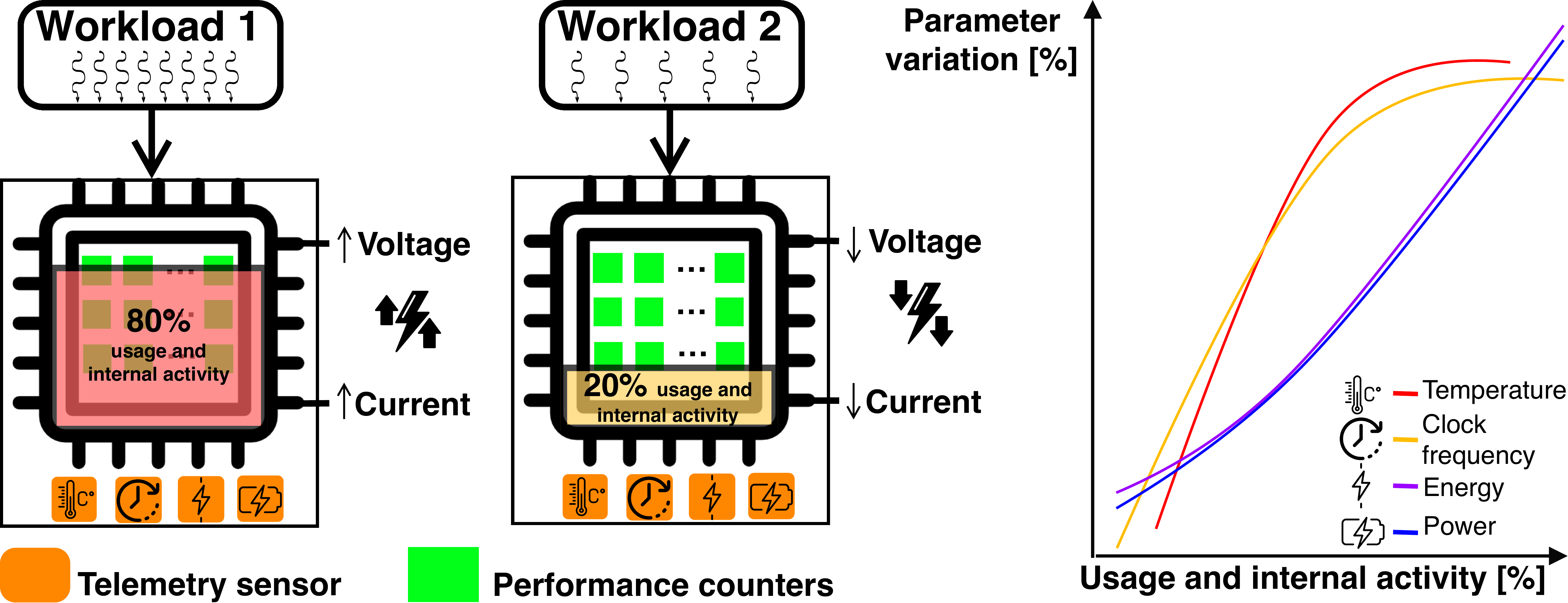}
    \caption{Two different workloads can have varying degrees of parallelism, which indicates a different number of threads executed simultaneously. A higher number of parallel threads requires greater resource usage, including computational resources, memory, and internal activity. As resource usage increases, performance counters that track internal activity and in-field telemetry sensors register greater variations in their recorded values. Additionally, the sensors embedded in the device show a larger variation in their readings.}
    
    % A general scheme of the GPU's internal activity for three workloads \todo{(each workload can be represented with a different number of threads, more threads means more usage, use symbols)}. For higher GPU resource usage, the Performance Counters increase the counter of the monitored event and, at the same time, the monitored telemetry parameters change due to the different usage, thus different switching activity, then different energy consumption. \todo{(unclear!!)} {\color{blue} (describe a general story of the 3 workloads and the effects of the parallelism, use of GPU resources, and then, their implications on the monitoring mechanisms, such as performance counters of in-field sensors, as well as other telemetry parameters.)
    % For higher GPU resource usage, the PCs (PCs) increase the recorded values and, at the same time, the sensors that collect telemetry data record higher variations of the monitored parameters (e.g., energy consumption).% \cite{ncu} \todo{[unclear the purpose of the reference, was it an adaptation from?, or was it copied from?]}
    \vspace{-6mm}
    \label{fig:corr}
\end{figure}
Different workloads affect devices in various ways, impacting hardware utilization, performance, power consumption, temperature, and frequency scaling. For instance, a computationally intensive application may force the 100\% device utilization, leading to increased power consumption and higher temperatures over time. Generally, the intensity of computation directly influences the thermal behavior of the device. Other factors, like frequency drops and energy consumption, can also indicate the stress level caused by the application. \,Dynamic\,Voltage\,and\,Frequency\,Scaling\,(DVFS) mechanisms can help reduce energy use and heat by lowering the device's operating frequency \cite{mei2017survey}.
% Every workload can have a distinct impact or signature on the devices, resulting in different hardware utilization, performance, energy/power consumption, temperature, and frequency scaling parameters. For example, a high computationally intensive application may force the 100\% device utilization, which can rapidly increase power consumption, which in the long run has an impact on the thermal behavior of the system (i.e, high temperatures). Typically, the computational intensity has direct influence with the thermal behavior of the device. However, other variables can also be used to estimate the level of stress induced by the application, such as frequency drop and energy consumption. In the first case, the \,Dynamic\,Voltage\,and\,Frequency\,Scaling (DVFS) mechanisms reduce energy consumption and heat generation by lowering the device's operating frequency \cite{mei2017survey}. 
Therefore, a significant frequency drop when executing a workload can be used as a stress indicator. On the other hand, energy consumption directly reflects the number of executed operations and the efficient use of GPU resources. 
% The workload complexity increase, implies more hardware resource usage, leading to higher power usage and energy expenditure. 
For example, more intensive computations or memory accesses raise the GPU’s energy consumption, making it a clear indicator of the stress level induced by the application \cite{tang2019impact}. These underlying operational dynamics, e.g., instruction counts, memory transactions, and execution efficiency, can be effectively captured through Performance Counters (PCs).
As a result, the stress induced by a workload can be defined as the strain placed on the hardware when it operates near its performance and operational limits.

    In this work, we propose a systematic evaluation strategy to combine internal events measurements with operative telemetry to effectively measure  the stress induced by different parallel workloads on GPUs. 
    The proposed solution combines workload profiling with the online monitoring of system telemetry parameters. 
    In particular, the profiling analysis describes the use of GPU resources by evaluating internal hardware events through the integrated and accessible Performance Counters in a GPU. 
    Moreover, the telemetry parameters provide the thermal, power, and clock frequency variables, resulting from executing parallel workloads that serve for monitoring the hardware internal state. % \todo{[that serve for ..... (computing performance?, power?, energy metrics?) why do you need them?]}
    
    % \todo{The reasons for using telemetry and performance counters are still not clear, the previous paragraph needs to clarify that by giving the arguments explaining the relation of all those variables and the induced "stress", and especially how this is different from the normal profiling mechanism }

% Our methodology integrates system-level telemetry monitoring with performance counter analysis, offering a dual perspective: \textit{i)} the device’s thermal and power response, and \textit{ii)} the application's efficiency in utilizing GPU computational units.

    In the experiments, we evaluated ten different applications that are drawn from three benchmark suites, including GPU-burn, a specialized state-of-the-art application to stress GPU devices, representative examples of CNNs and Rodinia Benchmarks. As GPU-burn is designed for stressing, we used its profiling results as the benchmark for GPU stress testing.
    We experimentally observed that CNN workloads can reach up to 63.19°C, which is nearly the maximum temperature of 64.97°C reached by GPU-burn which indicate moderate workload-induced stress. % \todo{[unclear this summary of results, what does it mean that 63.19 was reached by ML workloads? is it good or bad?]}
    However, CNNs issue only about 23\% of GPU-burn’s instruction rate due to the additional latency between instructions for memory access decreasing the overall induced stress of the GPU.
    % \todo{[due to??? what is the reason behind?, more use of the memories?, additional latency between the instructions?, is it good or bad? for the overall stress of the GPU?]}. 
    In contrast, other workloads generally lead to lower temperatures, reaching around 55.45°C. These workloads also exhibit a correspondingly lower instruction issue rate, peaking at approximately 22\% of GPU-burn’s activity due to the  low level of operation parallelism and the lower memory usage efficiency compared to GPU-burn. % This is due to the additional latency caused by both the low level of operation parallelism and the lower memory usage efficiency compared to GPU-burn, which reduces the overall stress on the system.
    % \todo{[This overview of the results can be improved by saying at the beginning that taking GPGPU-burn as the reference application, we experimentally observed that ML workloads provide less stress than the reference..... then providing the reasons why such behaviors occur, finally you can conclude saying in terms of the stress if this is good or not.]}

    The paper is organized as follows: \Cref{sec:eval} details the evaluation methodology used for the stress analysis of GPU applications. The experimental setup is described in \Cref{sec:expset}, followed by the presentation of the experimental results in \Cref{sec:results}. Finally, \Cref{sec:conclusions} outlines conclusions and future works.
\section{Stress analysis for GPU applications}
\label{sec:eval}

    This methodology characterizes the level of stress exerted on NVIDIA GPUs while running parallel workloads by monitoring telemetry data and PCs overtime.% \todo{[how?]} %\todo{where the CNNs in CUDA? I would say parallel workloads}, 
% with particular attention to identifying execution patterns that may compromise device reliability under sustained workloads. \todo{[until this point, it is unclear the main target of the methodology or what are you proposing, please improve this starting sentence, it is the one that helps the reader to understand the main point of the work (be direct with the description!)]}

As shown in \Cref{fig:corr} telemetry data, such as temperature and power consumption, can effectively measure stress levels in terms of energy required from the workload execution, PCs could be used to provide complementary and indirect indications about the computational stress in terms of internal hardware resource usage. The device usage (determined by the behavior of performance PCs) can be combined with thermal and system variables to better describe the stress induced by the workload during the online operation. 

Telemetry sensors, as shown in \Cref{fig:corr}, measure the system’s physical response to workload execution by capturing parameters. In contrast, PCs provide insights into internal hardware usage, offering complementary and indirect indications of computational stress. Specifically, a workload that demands significant resources triggers increased internal activity, including greater switching activity, which leads to higher energy consumption. This data from telemetry sensors can be combined with PCs data, along with thermal and system variables to better characterize the stress induced by the workload during online operations.

% \todo{[a scheme or a figure that correlates power, frequency, and performance countries information with stress would help to support the idea and would be easy for a reader to catch the main idea.]}

The evaluation proceeds in three main stages. First, a high-level profiling phase assesses how effectively the GPU's computational and memory subsystems are utilized during the application execution. As GPUs enable the collection of numerous PCs down to the warp level, measuring several internal events, it is essential to select a subset of these PCs to effectively estimate the stress levels caused by the running application. Based on the stress definition provided in \Cref{sec:intro}, the first profiling step allows us to identify the target PCs for measuring the stress induced by the selected workload. Next, the computational stress is characterized in detail by extracting hardware-level performance PCs, with a specific focus on the behavior of scheduling policy, memory subsystems, and compute units which represent the main functional GPU components for parallel computations. % \todo{[why these PCs, please explain!.. since /due to ....]}
Finally, telemetry data is collected to quantify the thermal stress imposed by the application. The joint use of PCs and telemetry data monitoring provides a comprehensive view of resource utilization and operating conditions during application run.

\subsection{High-level performance profiling}
\,NVIDIA's\,Compute\,Utility\,(NCU) \cite{ncu} is a profiling tool that facilitates roofline model analysis on GPU devices. We used this tool to identify the PCs that best represent the stress induced by a specific application. By employing roofline analysis, we evaluated how closely the applications operate to the hardware's performance limits. This allowed us to select the performance counters that better describe the efficiency of a specific workload in terms of hardware resource usage.

\begin{itemize}
    \item \textbf{Throughput} is calculated as the number of floating point operations executed per second (FLOP/s). The high throughput is reflected by the high number of executed arithmetic instructions within a time window. 
    % is calculated as the number of floating-point operations executed per second (FLOP/s) which indicate hoe. 
    \item \textbf{Arithmetic intensity} is defined as the ratio between the number of floating-point operations (FLOP) and the amount of memory traffic, expressed in FLOPs per byte. The high Arithmetic intensity can reflect the application bottleneck due to memory read and write instructions. % is defined as the ratio between the number of floating-point operations (FLOP) and the amount of memory traffic, expressed in FLOPs per byte. The memory traffic includes both reads and writes across different memory levels. % This metric indicates the computational density of the workload and helps determine whether its performance is limited by compute capacity or memory bandwidth. 

    % \todo{[no idea what is the purpose of defining throughput and intensity, for an expert in the field, both are obvious, it would be clearer to mention that you compute the roofline model and cite the paper, that will be enough. Then, in the discussion, you can analyze the results based on the definitions.] [Another alternative if you decide to keep both definitions is to mention and indicate that throughput and intensity are direct indicators of stress in an app or device.]}
\end{itemize}

% Using these metrics, \gls{ncu} generates a roofline model that visualizes the application's performance relative to architectural limits. 
% This study provides a high-level characterization of the stress imposed on the GPU by GPU-burn, and serves as a baseline for comparison against other evaluated workloads.

\subsection{Fine-grain Performance Measurement}
\begin{table}[]
\caption{Detailed description of the selected performance counters % \todo{this table can be compressed!} % \todo{use a short name in one column, and then the explanation in another colum}
}
\begin{tabular}{l|l|l}
\hline
\textit{\textbf{Group}}                                                & \textit{\textbf{CUPTI PCs}} & \textit{\textbf{Description}}                                                                                                            \\ \hline
\multirow{7}{*}{\textbf{Throughput}}                                   & \begin{tabular}[c]{@{}l@{}}ALU \\ instructions\end{tabular}                               & \begin{tabular}[c]{@{}l@{}}Executed operations in \\ Arithmetic Logic Units per SM\end{tabular}                                           \\ \cline{2-3} 
                                                                       & \begin{tabular}[c]{@{}l@{}}FP16 \\ instructions\end{tabular}                              & \begin{tabular}[c]{@{}l@{}}Executed operations \\ between FP16 per SM\end{tabular}                                  \\ \cline{2-3} 
                                                                       & \begin{tabular}[c]{@{}l@{}}FP64 \\ instructions\end{tabular}                              & \begin{tabular}[c]{@{}l@{}}Executed operations \\ between FP64 per SM\end{tabular}                                  \\ \cline{2-3} 
                                                                       & \begin{tabular}[c]{@{}l@{}}DMMA \\ instructions\end{tabular}                              & \begin{tabular}[c]{@{}l@{}}Executed Dot Product operations \\from Tensor Core Unit per SM\end{tabular}                               \\ \cline{2-3} 
                                                                       & \begin{tabular}[c]{@{}l@{}}HMMA \\ instructions\end{tabular}                              & \begin{tabular}[c]{@{}l@{}}Executed operations between FP16 \\ from Tensor Core Units per SM\end{tabular}                             \\ \cline{2-3} 
                                                                       & \begin{tabular}[c]{@{}l@{}}IMMA \\ instructions\end{tabular}                              & \begin{tabular}[c]{@{}l@{}}Executed operations between Integers \\from Tensor Core Units per SM\end{tabular}                         \\ \cline{2-3} 
                                                                       & \begin{tabular}[c]{@{}l@{}}XU \\ instructions\end{tabular}                                & \begin{tabular}[c]{@{}l@{}}Executed special operations\\(e.g., sin cos, log) per SM\end{tabular}                                   \\ \hline
\textbf{\begin{tabular}[c]{@{}l@{}}Issued\\ instructions\end{tabular}} & \begin{tabular}[c]{@{}l@{}}Issued \\ instructions\end{tabular}                            & \begin{tabular}[c]{@{}l@{}}Issued instruction per SM SubPartition\end{tabular}                                              \\ \hline
\multirow{3}{*}{\textbf{Stall events}}                                 & \begin{tabular}[c]{@{}l@{}}Memory \\ stall\end{tabular}                                   & \begin{tabular}[c]{@{}l@{}}Stall due to memory dependency.\end{tabular}                                                               \\ \cline{2-3} 
                                                                       & \begin{tabular}[c]{@{}l@{}}Scheduler \\ stall\end{tabular}                                & \begin{tabular}[c]{@{}l@{}}Stall due to divergences \\ or kernels syncronization \\ issues (e.g., divergences or barriers).\end{tabular} \\ \cline{2-3} 
                                                                       & \begin{tabular}[c]{@{}l@{}}Throttle \\ stall\end{tabular}                                 & \begin{tabular}[c]{@{}l@{}}Stall due to throttle mechanism\\activated for lack of available resources.\end{tabular}      \\ \hline
\end{tabular}
\vspace{-5mm}
\label{tab:PC_description_pt2}
\end{table}
% After establishing a high-level performance baseline, a fine-grain performance analysis is conducted to compare the internal stress behavior of all evaluated applications. \todo{It is not explained the reasons why we need to establish a baseline? } This stage focuses on 
% As a second step, a deeper view on the resource usage, is provided
% quantifying how each workload interacts with architectural components, providing detailed insight into the computational and memory pressure exerted on the device. \todo{Too ambiguous sentence}
% On the one hand, the Roofline model
As the second step of our evaluation, PCs are collected through the CUDA Profiling Tools Interface (CUPTI) \cite{cupti} to monitor the internal GPU activity. CUPTI grants access to PCs that monitor the systems activity such as \glspl{sm}, schedulers, caches (L1 and L2), and controllers (e.g., DRAM). To collect these metrics, one of the sample scripts included in the CUPTI toolkit—specifically the profiling\_injection script—was adapted to extract the target PCs. % \todo{why those counter can give the level of analysis required, how they can be used for estimating stress? wouldn't it betterto mention you collected them all, and later using a data evalaution you selected the ones that allowed to differentiate the behaviour of each application???}

\begin{table}[]
\caption{Metric derived from the Performance Counters sampling.}
\begin{tabular}{c|c|c}
\hline
\textit{\textbf{\begin{tabular}[c]{@{}c@{}}Metric \\ name\end{tabular}}} & \textit{\textbf{Formula}}                                                              & \textit{\textbf{Description}}                                                                                                                             \\ \hline
\textbf{\begin{tabular}[c]{@{}c@{}}SM \\ busy\\ rate\end{tabular}}       & $\frac{\frac{\#Arithmetic\_instructions}{Warp\_size}}{SM\_busy\_rate_{Pk}} \times 100$ & \begin{tabular}[c]{@{}c@{}} Instructions executed \\ per cycle per SM \\ w.r.t. its HW peak.\end{tabular}                          \\ \hline
\textbf{AII}                                                             & $\frac{\frac{\#Issued\_instructions}{\# Active\_cycles}}{AII_{Pk}} \times 100$         & \begin{tabular}[c]{@{}c@{}}Requested instructions \\ but not executed \\  per active cycles per \\ SM w.r.t. its HW peak.\end{tabular}  \\ \hline
\textbf{S$_{act}$}                                                       & $ \frac{S_{throt} + S_{cont} + S_{mem}}{S_{TOT}}\times 100 $                           & \begin{tabular}[c]{@{}c@{}}Percentage of stalls \\ related to device activity \\ w.r.t. all stall reasons.\end{tabular} \\ \hline
\end{tabular}
\vspace{-5.5mm}
\label{tab:metrics_pt2}
\end{table}

    GPUs enable the collection of numerous PCs down to the warp level, measuring several internal events; therefore, it is essential to select a subset of these PCs to effectively estimate the stress levels caused by the running application.
    Following the profiling guidelines provided by NVIDIA \cite{profiling}, we report in \Cref{tab:PC_description_pt2} 11 of PCs that capture three different aspects of the workload: %\todo{[how do you select them and based on what? Did you do an exploration to determine the most representative ones, or did you use any other strategy? Here, it is important to describe.]} 
    \textit{i) throughput}, focusing solely on instructions that carry out mathematical operations; \textit{ii) issued instructions}; and \textit{iii) stalls} caused by available resources, such as memory and computational units, as well as those caused by the scheduler. For each category, we computed the metrics outlined in \Cref{tab:metrics_pt2}. These guidelines were interpreted with the specific goal of identifying the architectural features and usage patterns (e.g., usage of L1 or L2 cache memory) that contribute most significantly to inducing a sustained computational workload, thereby maximizing stress on the GPU
%\todo{Which are such architectural features and usage patterns???}
. We focused on metrics capable of capturing GPU stress from two perspectives: throughput, represented by the \textit{SM busy rate} as it reflects computational stress by indicating how intensively the GPU's execution units are utilized, \textit{\,Active\,Issued\,Instructions\,(AII)} which indicates the stress on the instruction pipeline by quantifying how many instructions are ready to execute; and the percentage of stall events due to internal activity, denoted as $S_{act}$ which reveal bottlenecks in the computations due to stalled or blocked execution from activity perspective. % \todo{Why the busy rate and the stalls can be associated to GPU stress?}

% \todo{It was not mentioned anywhere that a GPU have thousands of PCs, measuring quite a lot of internal events, therefore it is fundamental to select a subset of such PCs that can be used for estimate the stress levels induced by the executed application. ...}

\subsection{Telemetry monitoring}
To complete the stress evaluation, telemetry-based measurements are incorporated to account for the physical response of the device. While performance PCs capture architectural-level activity, telemetry metrics reflect how internal stress translates into observable thermal and power dynamics.

NVIDIA Management Library (NVML) \cite{nvml} enables system-level access to internal sensor data with a configurable sampling period. NVML provides real-time insights into the GPU's operating conditions, allowing us to profile stress by tracking key physical (e.g., Temperature) and dynamic indicators (e.g., Clock Frequency). We selected the following for their relevance to system stress:
\begin{itemize}
    \item \textit{\,Energy\,Consumption\,(E)}: reflects both the duration and intensity of a workload, making it a reliable indicator of long-term physical degradation mechanisms. As the time integral of power, total energy captures the number of executed operations and the efficient use of GPU resources. To quantify energy accumulation, we calculated the difference between final and initial energy readings and normalize by the experiment's duration (\textit{E}) to compare workloads of different lengths.
    
    \item \textit{Temperature\,(Temp)}: As the GPU operates, the more internal switching activity, the more energy consumed and the more heat dissipated. This heat accumulation reflects the device's internal stress and thermal load. % \todo{[unclear description of this parameter, be clear and simple]}
    
    \item \textit{SM\,Clock\,Frequency\,(CF)}: The SM clock frequency reflects the operative state of the GPU's SM. Thermal stress triggers DVFS, a protective mechanism that reduces performance to maximum self-heating figures. As it is not possible to measure voltage at the system level, the average variation (\textit{$\Delta$ CF}) w.r.t. the peak CF experienced during the application run provide an indirect but informative signal of dynamic adaptation to the induced stress.
\end{itemize}

To further characterize thermal behavior, we analyzed the temperature profile over time, which typically follows an exponential trend, accordint to the lumped-capacitance thermal model \cite{tegenaw2019comparison}. %, the temperature evolution can be approximated with \Cref{eq:thermal}.
% \begin{equation}
%     T(t) = T_\infty - (T_\infty - T_0) e^{-t/\tau}
%     \label{eq:thermal}
% \end{equation}
% where $T_0$ is the initial temperature, $T_{\infty}$ the steady-state temperature, $t$ the time, and $\tau$  the system’s thermal time constant. 
This model allows us to define two key metrics that characterize the stress imposed on the system: \textit{i)} $t_r$ the time required for the temperature to reach a steady-state condition computed as in \cite{truong2020selective}, and the lower $t_r$, the faster the system's thermal reaction, the higher the stress induced in a reduced amount of time, and \textit{ii)} $T_{\infty}$ the average temperature recorded during the steady-state interval. DVFS mechanisms stabilize the system by dynamically adjusting clock frequency and voltage to optimize performance and power efficiency. $T_{\infty}$ represents the thermal equilibrium point; higher $T_{\infty}$ values indicate that this equilibrium is achieved under greater and sustained energy consumption, reflecting increased physical stress on the GPU over time.% \todo{Not clear how this variable can be used for stress measurement, provide more details regarding the interpretation in terms of stress measurement.}

\section{Experimental setup}
\label{sec:expset}
% \todo{Descrivi i benchmark e cosa differisce GPU-burn dagli altri}
% \input{tables/experimental_setup}
    % For our evaluation, we selected a total of ten \gls{gpu} applications spanning different computational domains. %The \gls{gpu}-burn application, which repeatedly performs matrix multiplications, is designed to stress the \gls{gpu} and detect potential data corruption by comparing results against a precomputed ground truth \cite{defour2013gpuburn}. 
    % In addition to \gls{gpu}-burn\todo{Why "in addition", gpu-burn is one of the applications}
    Our application set includes ten programs, grouped into three categories: \textit{i)} GPU-Burn, primarily used for stress testing, which executes multiple instances of a matrix multiplication kernel to push the GPU to its computational limits; and \textit{ii)} inference workloads from four representative neural networks for image classification—LeNet-5 on MNIST, and MnasNet, MobileNetV2, and ResNet18 on CIFAR-10—all of which invoke a variety of CUDA kernels to perform their respective tasks. As the third category, we incorporated five CUDA applications from the Rodinia benchmark suite: Back Propagation, Gaussian Elimination, Hotspot, Needleman-Wunsch, and Streamcluster. These benchmarks are widely adopted in research to evaluate the efficiency and scalability of heterogeneous architectures and parallel programming techniques.% \todo{[since you are focusing on stress, it would be good to have additional information regarding the application, total execution times, and number of parallel instructions. A table can help to summarize this info.]}

    % We configured the applications to be run for 5 minutes (the maximum time to reach the steady-state temperature across all the applications) 
    % \todo{Potentially, you can do five runs of each experiment. If you do so, report mean and standard deviation for all the numbers reported in the results}
    We configured the applications to be run during 5 minutes (we empirically found that this is the maximum time to reach the steady-state temperature across all the applications) with a memory occupancy of 50\% to balance high computational load with sufficient headroom to accommodate the overhead introduced by CUPTI instrumentation. 
    % Specifically, \gls{gpu}-burn allows to specify the duration and the memory occupancy w.r.t. the device's available memory. \todo{Why is it important to point out the gpu-burn configurations? [I have the same question!!!!!!]}
    
    We repeated the execution of Rodinia benchmarks for
    % we retained the original parameter sets from \cite{che2009rodinia} \todo{[which ones?]} and 
    a sufficient number of iterations to match the reference duration. The number of iterations for Rodinia applications ranges from 1,500 for Gaussian Elimination to 45,000 for Back propagation due to the various parallelization and memory access patterns. 
    
    In the case of the ML workloads, we adjusted the batch size to match the target memory footprint and repeated the inference for the required number of iterations to reach the desired execution time. This required the Batch sizes to go from 4,096 for Resnet18 to 10,000 for LeNet5, repeating the inference for around 100 iterations % \todo{start the paragraph mentioning that you configure all application to be run during 5minutes, why 5 min?. then you can proceed with the details and explanations that support the execution during the selected amount of time.

Telemetry parameters are collected every second during application execution to balance temporal resolution and data volume. %\todo{[this sentence can be inverted as: To provide a balance between temporal resolution and data volume, the telemetry parameters are collected with a 1-second interval during each application's execution.]} 
Simultaneously, PCs were sampled after each kernel invocation. % to capture detailed insights into the evolution of \gls{gpu} activity.
The application monitoring was performed on a laptop MSI Cyborg 15 A13VF with an Intel Core i7-13620 CPU with 20 cores, 16 GB of RAM, and equipped with one NVIDIA GPU RTX 4060 with CUDA version 12.6.

% \todo{[until here was not clear the type of GPU you used for the experiments (what is the reference?), the CUDA version or any other relevant configuration of the experiments.!!!!]}
\section{Experimental results}
\label{sec:results}

\subsection{Roofline analysis}
\begin{figure}[tb]
    \centering
    \includegraphics[width=0.85\linewidth]{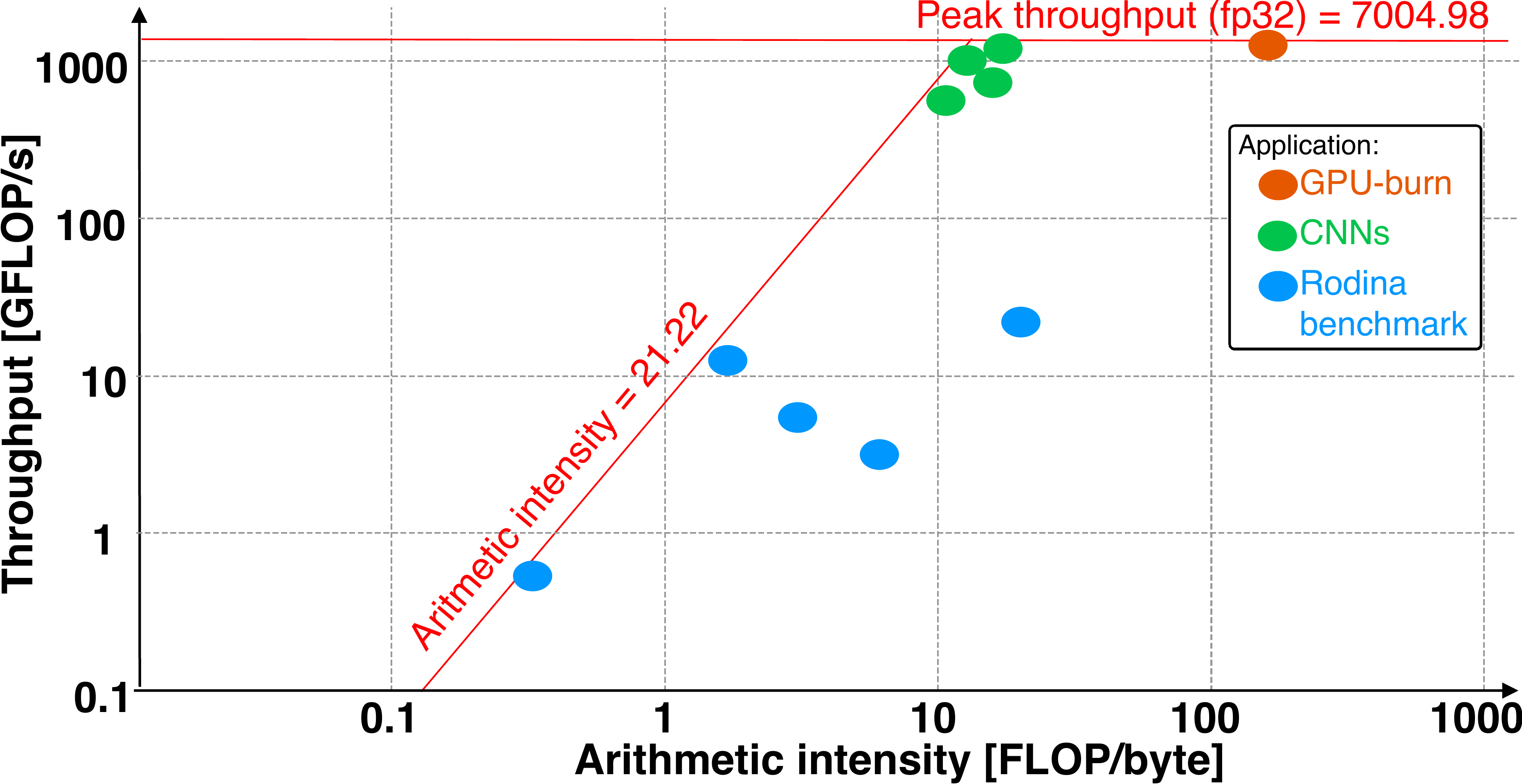}
    \caption{Roofline model analysis for the evaluated benchmarks.% \todo{Attento al bound sull'aritmetic intensity}. % \cite{ncu} \todo{[unclear the purpose of the reference, was it an adaptation from?, or was it copied from?]}
    }
    \label{fig:roof}
    \vspace{-6mm}
\end{figure}

% The Roofline model provides a high-level framework for evaluating application performance by relating computational throughput (FLOP/s) to arithmetic intensity (FLOP/byte) \cite{ncu}. Specifically, the aritmetic intensity is defined as the ratio between the total number of floating-point operations and the number of bytes transferred across the memory hierarchy. 
\Cref{fig:roof} reports the obtained Roofline analysis for the evaluated benchmarks. The theoretical performance ceilings are plotted as red lines and represent upper bounds imposed by the hardware’s architectural constraints. 
% \todo{It is important to indicate somewhere across the paper that the theoretical performance ceilings are the maximum possible performance delivered by the GPU; not possible to go further than that.} 
Specifically, the target GPU in this study supports an ideal computational throughput of 6,450 GFLOP/s and a maximum global memory bandwidth of 25.22 GB/s. Overall, 3 main clusters of workloads can be identified corresponding to GPU-burn, CNNs, and Rodinia benchmark workloads. Each cluster shows different levels of resource usage efficiency, with GPU-burn almost reaching the peak hardware performance.

Profiling results demonstrate that GPU-burn operates with an arithmetic intensity of 181.25 FLOP/byte and a throughput of 4,019 GFLOP/s, which indicates highly efficient reuse of data and near-maximum utilization of the computational units. 
Indeed, GPU-burn is the reference state-of-the-art benchmark for GPU stress testing. Thus, for the purpose of this work, the GPU-burn Roofline profile serves as a reference for the operational conditions for maximum stress on a GPU. 

% Among all tested workloads, GPU-burn consistently imposes the highest sustained stress on both the memory and compute subsystems. 

%\todo{How do you know that GPU burn induces the highest stress based on the roofline model?, I think here we associate the previous knowledge based on the state-of art, as GPU-burn has been developed for that purpose, therefore understanding the roofline model for GPUburn, helps us to set a reference operational conditions for maximum stress induced on the GPU device.} % Its position on the Roofline makes it an ideal reference benchmark for identifying the upper bounds of stress-inducing behavior in real-world applications.

CNN workloads show diverse behavior depending on model complexity, but generally operate below the GPU-burn performance.
Mobilenet, MnasNet exhibit arithmetic intensities ranging from 11.22 to 12.42 FLOP/byte (indicating application bottleneck due to memory usage bounding) and throughput of 802.64 to 1,882.94 GFLOP/s, indicating moderate but not saturating usage of the GPU. Lenet shows an arithmetic intensity of 19.92 FLOP/byte but with a throughput of 967.70 GFLOP/s.  % \todo{[general question: the figure 1 clearly shows 2 CNNs with memory bounding, which of these CNNs are affected by this behavior?]}
ResNet stands out with an intensity of 25.24 FLOP/byte and a throughput of 2,306.32 GFLOP/s, approaching the compute-bound region. 
Compared to GPU burn, CNNs create significantly lower stress levels, although ResNet approaches high-stress levels. 
CNNs use convolution, which involves matrix multiplications like GPU-burn workloads. However, unlike GPU-burn, CNNs perform additional intermediate operations such as pooling and activation functions while processing smaller matrices. These extra operations use fewer resources and lessen GPU stress by reducing arithmetic instructions and introducing memory-bound bottlenecks. % \todo{Missing a conclusive sentence to this paragraph.}

% \todo{CNNs are well known for being highly computationally intensive, but why are they not as computationally intensive as GPU-burn? } % Overall, CNNs demonstrate better efficiency than traditional HPCs workloads, yet still leave a gap from the GPU’s theoretical peak. 

Applications from the Rodinia suite continue to populate the lower left region of the Roofline plot, indicating limited arithmetic intensity and suboptimal computational throughput compared to both CNNs and GPU-burn. Needleman-Wunsch shows the weakest performance, with an arithmetic intensity of just 0.57 FLOP/byte and a throughput of 0.41 GFLOP/s. Stream Cluster and Hotspot demonstrate modest improvements, reaching 5.27 and 2.34 FLOP/byte in intensity and 7.51 and 11.69 GFLOP/s in throughput, respectively—yet still well below the GPU’s peak capabilities. Back Propagation stands out slightly, achieving a relatively high arithmetic intensity of 29.76 FLOP/byte and a throughput of 33.34 GFLOP/s, indicating a more compute-focused workload. In comparison to CNNs and GPU-burn, Rodinia applications are the least demanding, revealing a lower degree of computational intensity due to the higher amount of executed arithmetic instructions and of stalls due to memory bandwidth limitations. %\todo{[or computational intensity?, at this point you can only guess, the performance counter can guide you to identify the reasons for the low throughput]} that limits, in principle, their ability to stress GPU resources in a sustained fashion. 

% \todo{Which kind of deficiencies? degree of parallelism, algorithm? or low performance due ot implmentation?}

% Applications from the Rodinia suite generally occupy the lower left region of the Roofline plot, indicating limited computational throughput and low arithmetic intensity. Most benchmarks are memory-bound: Needleman-Wunsch and Stream Cluster achieve low intensities (0.27 and 0.43 FLOP/byte) and minimal throughput (2.31 and 27.77 GFLOP/s), while Back Propagation and Hotspot perform slightly better—with arithmetic intensities of 8.52 and 5.27 FLOP/byte and throughputs of 5.33 and 7.51 GFLOP/s, respectively—but still remain far from optimal use of the hardware. Gaussian is the only Rodinia workload to reach a higher intensity of 29.76 FLOP/byte, though its throughput remains low at 33.34 GFLOP/s. In comparison to CNNs and GPU-burn, Rodinia applications are the least demanding, revealing inefficiencies that limit their ability to stress GPU resources in a sustained fashion.

\subsection{Performance stress indicators}
\begin{figure*}[tb]
    \centering
    \includegraphics[width=0.85\linewidth]{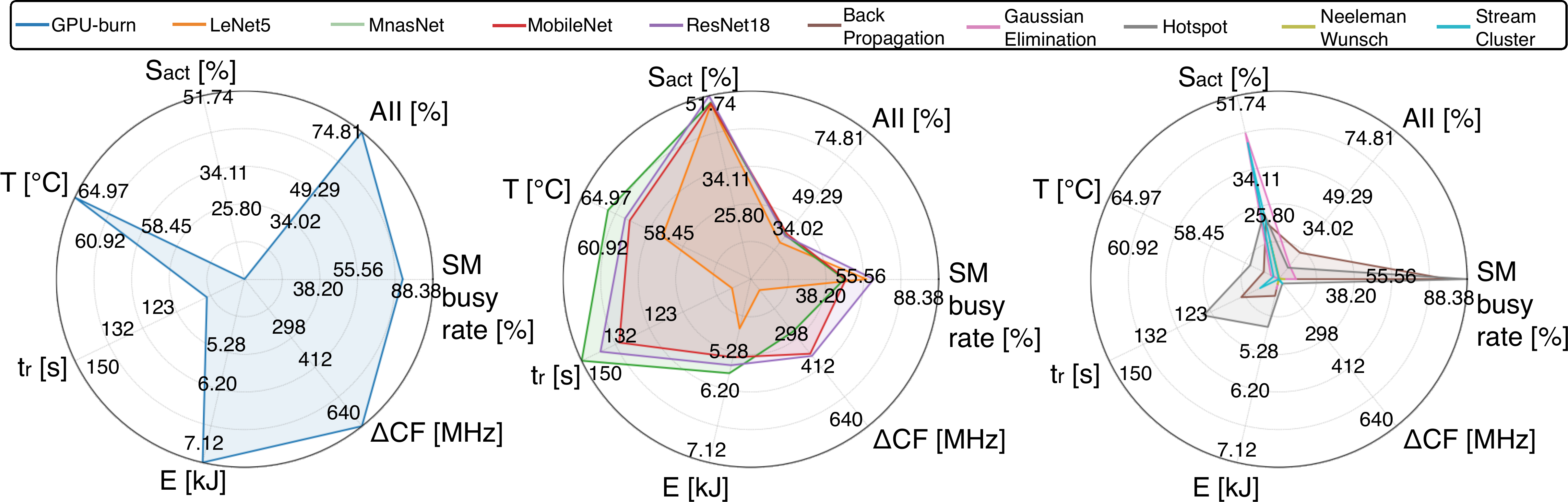}
    \caption{Radar chart showing the telemetry and PCs data for the evaluated workloads (i.e., GPU-burn\cite{defour2013gpuburn}, CNNs \cite{marcel2010torchvision} and Rodinia benchmarks \cite{che2009rodinia}.}
    \label{fig:radar}
    \vspace{-4mm}
\end{figure*}

\Cref{fig:radar} provides a summary of architectural and physical stress indicators across evaluated GPU workloads. Each axis reports the performance and telemetry metrics, including SM utilization (SM busy rate), instruction activity (AII), stall ratio ($S_{act}$), thermal-dependent behavior (stationary temperature $T_{\infty}$, thermal response time ($t_r$), E and $\Delta$ CF. A higher level of stress is indicated by a higher SM busy rate, AII, $T_\infty$, E and $\Delta$ CF but a lower $t_r$ and $S_{act}$.
% Based on the results, we used color encoding to rank application stress: red (high), orange (medium), and yellow (low).

Among the evaluated workloads, CNN-based models and Rodinia benchmarks exhibit significantly lower stress--in terms of internal activity--on the GPU’s computational architecture compared to GPU-burn.
% \todo{stress in terms of? temperature, internal GPU activity? them all together? Is there any variable that better describes the stress???} 
While GPU-burn achieves an SM busy rate of 74.82\%, CNNs range between 45.74\% (MnasNet) and 59.04\% (ResNet18). This reduction is due to the mixed nature of CNN operations, which alternate between MAC-dominated phases and others, e.g., reductions or comparisons, that shift focus from computation to memory access. These variations depend on the specific layer type (e.g., convolution, activation, pooling). Instruction activity reflects this memory-bound behavior: while GPU-burn reaches an AII of 74.81\%, CNNs report significantly lower values, between 21\% and 25\%. In fact, stall due to internal activity ($S_{act}$) is 40\% higher in CNNs, w.r.t. just 9.16\% for GPU-burn. At least 35\% of stalls in each CNN are directly caused by memory operations. %\todo{Which kind of delays? stalls? or physical silicon delays?}

The remaining applications span various computational domains and show diverse architectural stress profiles. Some, like Hotspot and Back Propagation, achieve very high SM utilization (88.38\% and 79.61\%, respectively), exceeding that of GPU-burn. However, this does not directly translate into higher instruction activity: their AII values remain well below GPU-burn (e.g., 9.00\% for Hotspot, 16.30\% for Back Propagation), indicating that SMs are often active without issuing instructions at the same rate. This is largely due to synchronization events—accounting for about 15\% of stall reasons—that block compute units without maintaining high throughput. Applications like Gaussian Elimination, Needleman-Wunsch, and Stream Cluster show extremely low SM utilization (between 3.48\% and 11.21\%) and minimal instruction throughput (AII $<$ 10\%). Their $S_{act}$ values, however, remain non-negligible, suggesting that internal stalls persist even under light computational loads. This behavior likely stems from inefficient GPU utilization, memory bottlenecks, or poor parallelism—particularly in legacy or irregular algorithms not optimized for GPU architectures.

Overall, GPU-burn demonstrates very high utilization of GPU resources, achieving a 74.82\% SM busy rate and 9.162\% stalls. Other applications achieve a busy rate of 79.61\% in Back Propagation and 88.38\% in Hotspot, but the induced stress is limited by their low AII values (16.30\% and 9.00\%) and high stall percentages. The remaining applications reach a maximum of 60\% of the SM busy rate of GPU-burn and exhibit at least a 50\% increase in stalls.

\subsection{Performance Metrics impact on Thermal and Dynamic Stress}
CNN applications induce moderate physical stress compared to GPU-burn, as confirmed by telemetry data. The steady-state temperatures—ranging from 59.5 °C (LeNet5) to 63.2 °C (MnasNet)—remain just below GPU-burn’s peak (64.97°C), reflecting sustained but not saturating hardware activity. These thermal levels align with their architectural profile: while CNNs maintain moderate SM utilization (45–59\%), their instruction activity is significantly lower (AII 21–25\%) due to the alternation of compute-intensive and memory-dominated phases. This leads to substantial internal stalls ($S_{act}$ 13–15\%) and ultimately lower switching activity, which reduces overall power draw (5.44–6.00kJ) compared to GPU-burn (7.12kJ). The limited clock frequency variation ($\Delta$CF of 120–420MHz, vs. 740MHz for GPU-burn) further confirms that DVFS mechanisms are less frequently triggered, implying stable thermal behavior.

Rodinia benchmarks show more heterogeneous patterns, driven by varied computational efficiency. Back Propagation and Hotspot reach intermediate thermal stress levels (T$_\infty>$54°C, E$=$5.4kJ), yet exhibit low AII (9–16\%), revealing inefficient instruction issue despite high SM utilization ($>$79\%). Their moderate stall values (11–14\%) and small $\Delta$CF (70–90MHz) suggest that the thermal buildup is caused more by sustained occupancy than by dynamic throughput. Conversely, applications such as Gaussian Elimination, Needleman-Wunsch, and Stream Cluster generate minimal thermal stress, with low temperatures (53–54°C), energy (4.82–4.94kJ), and almost flat clock profiles—consistently explained by their extremely low SM activity (3–11\%) and instruction throughput (AII $<$ 10\%), likely due to poor parallelism or memory bottlenecks.

In conclusion, the joint analysis of telemetry data and PCs enables a comprehensive assessment of application-induced GPU stress. Isolated telemetry metrics provides direct indicators of physical stress—such as temperature, and energy consumption. On the other hand, PCs, if simultaneously monitored, can offer indirect insights by quantifying how intensively an application utilizes GPU resources. This approach identifies benchmarks that increase stress and operating temperatures, showing how applications with high parallelism affect GPU stress. Combining these insights is essential for predicting long-term reliability risks and informing testing and workload deployment decisions. This includes assessing execution times and determining the necessary duration for a stress test to uncover fault effects.
% \todo{[missing a general closing paragraph that highlights that the combination of temeletry, performance and roofline profiles allows the prediction and identification of benchmarks that are highly prone to stress and GPU and impact their operative temperatures, which indicates that at the long term, aging effects might start to arise.]}

% \todo{[what is the purpose of the work, is it intended to serve as a guide for the selection of clever and minimal parameters in the computation of stress figures?]}

% \todo{It would be great to add some insights about the usage of telemetry and performance metrics as stress measurements, and how they can contribute to defining the periodic test using GPU-burn in a real deployment, at least using a qualitative description}

% telemetry data reinforce the characterization of GPU-burn as a worst-case stressor
%\todo{I think from the state of art we all know that GPU-burn exhibits more stress, The main focus should be how you characterize the stress using certain set of of variables monitored in field. Also, consider the fact that there might be variables that are not necessarily useful for this purpose, the conclusion and discussion should provide some insights about that},
% \input{sections/04_Discussion}
\section{Conclusion and future works}
\label{sec:conclusions}
In this work, we presented a methodology for characterizing the stress induced by GPU-based applications through the combined analysis of GPU telemetry and PCs. 
By evaluating ten representative applications, we demonstrated how different workloads impose varying levels of thermal and computational stress on the GPU, thus exacerbating the need to identify relevant parameters to estimate stress levels through the combination of the measured parameters.

% \todo{[and?.... was it the main purpose?, or the main purpose was to support the identification of stress levels in apps through the combination of the measured parameters?]}.

% GPU-burn is a reference stress-testing workload for GPUs \cite{defour2013gpuburn}. 
As GPU-burn was devised to induce stress in GPUs, it reaches the highest steady-state temperature of 64.97 °C and consuming 7.12 kJ. These physical indicators align with intense hardware usage, as shown by its SM busy rate of 74.82\% and AII of 74.81\%, and are further supported by the largest observed DVFS variation ($\Delta$CF = 740 MHz). In contrast, our results show that CNN workloads exhibit moderate stress levels, with AII values of 21–25\%, and $\Delta$CF not exceeding 420 MHz. 
Rodinia benchmarks remain below these thresholds due to inefficient resource utilization, low parallelism, and memory-bound behavior, which result in both lower performance and thermal impact.
These findings demonstrate that telemetry and performance metrics are strongly correlated: telemetry data provide a direct view of the physical stress, while PCs capture the computational behavior that causes it. 
Together, they offer a reliable way to assess workload-induced stress and identify applications that, if run repeatedly, may accelerate aging and increase the risk of hardware degradation over time.

    In future works, we plan to compare the stress impacts experienced by new and thermally compromised GPUs in data centers, as well as to expand our benchmark suite to include additional workloads.
    % \todo{Use GPU for data centers to validate the scalability of our evaluation methodology and experimental setup}

% \todo{It is not yet clear how a workload can be characterized in terms of stress. The paper gives a series of metrics and variables, but there are no clear ideas about the role of such selected metrics.  What is the core idea of the paper? Is it a paper on selecting the best variables for measuring stress? Are we proposing a stress characterization based on some arbitrary selected metrics?, Define the path and stick to it because, so far, the paper is too ambiguous as the ideas and contributions are not clear, and the conclusions and results are difficult to assess.}

\section*{Acknowledgments}
This work has been supported by the National Resilience and Recovery Plan (PNRR) through the National Center for HPCs, Big Data and Quantum Computing.

\bibliographystyle{IEEEtran}
\bibliography{reb}

\end{document}